\documentclass[a4paper,11pt,preprintnumbers]{article}
\usepackage{pos}

\usepackage{graphicx} 
\usepackage{amsmath}
\usepackage{physics}
\usepackage{cleveref}
\usepackage{xcolor}
\usepackage{multirow}
\usepackage{todonotes}
\usepackage{lineno}

\title{Split-even approach to the rare kaon decay $K \to \pi \ell^+ \ell^-$}

\author*[a]{Raoul Hodgson}
\author[b]{Vera G\"ulpers}
\author[b]{Ryan Hill}
\author[b]{Antonin Portelli}

\affiliation[a]{Deutsches Elektronen-Synchrotron DESY, Platanenallee 6, 15738 Zeuthen, Germany}
\affiliation[b]{Higgs Centre for Theoretical Physics, School of Physics and Astronomy, The University of Edinburgh, Edinburgh EH9 3FD, United Kingdom}

\emailAdd{raoul.hodgson@desy.de}

\abstract{In recent years the rare kaon decay has been computed directly at the physical point. However, this calculation is currently limited by stochastic noise stemming from a light and charm quark loop GIM subtraction. The split-even approach is an alternative estimator for such loop differences, and has shown a large variance reduction in certain quantities. We present an investigation into the use of the split-even estimator in the calculation of the rare kaon decay.}

\FullConference{The 41st International Symposium on Lattice Field Theory (LATTICE2024)\\
 28 July - 3 August 2024\\
Liverpool, UK\\}

\begin{document}

\begin{flushright}
DESY-25-016
\end{flushright}
\vspace{-1.3cm}

\maketitle

\section{Introduction}
\subsection{Phenomenology}
The rare FCNC decay of kaons into pions and two leptons can be used to search for new physics beyond the SM. One possible channel for such decays is the charged lepton mode $K \to \pi \ell^+ \ell^-$. This decay is dominated by the long distance process $K \to \pi \gamma^\ast \to \pi \ell^+ \ell^-$ which pair produces the leptons from an intermediate virtual photon. Therefore the goal is to compute the $K \to \pi \gamma^\ast$ amplitude which is given by
\begin{align}
\label{eq:Amplitude_Def}
    \mathcal{A}_\mu^i & = \int d^4 x \bra{\pi^i,\boldsymbol{p}} T\{ H_W(x) J_\mu(0) \} \ket{K^i,\boldsymbol{k}} \,,
\end{align}
where the label $i=+,S$ indicates the charged or neutral kaon decay processes $K^+ \to \pi^+ \gamma^\ast$ and $K_S \to \pi^0 \gamma^\ast$ respectively. $\boldsymbol{k}$ and $\boldsymbol{p}$ are the kaon and pion 3-momenta respectively, $J_\mu$ is the electromagnetic current, and $H_W$ is the effective $s\to d$ quark weak Hamiltonian described by \cite{RevModPhys.68.1125}
\begin{align}
\label{eq:Weak_Hamil}
    H_W = \frac{G_F}{\sqrt{2}} V_{us}^\ast V_{ud} \left[ C_1 (Q_1^u-Q_1^c) + C_2 (Q_2^u-Q_2^c) + \dots \right] \,.
\end{align}
$C_i$ are the Wilson coefficients corresponding to the the 4-quark operators $Q_{1,2}^q$ operators given by
\begin{align}
    Q_1^q = & [\bar{d} \gamma_\mu (1-\gamma_5) s] [\bar{q} \gamma^\mu (1-\gamma_5) q] \\
    Q_2^q = & [\bar{d} \gamma_\mu (1-\gamma_5) q] [\bar{q} \gamma^\mu (1-\gamma_5) s] \,.
\end{align}
The Wilson coefficients of $C_{i>2}$ are suppressed compared to $C_{1,2}$ \cite{ISIDORI200675} and therefore to the desired level of precision in the near future, only these two operators are required.
The operator differences $Q_i^u-Q_i^c$ present in \cref{eq:Weak_Hamil} are a manifestation of the GIM cancellation when renormalizing to a scale above the charm mass.

The amplitude in \cref{eq:Amplitude_Def} can be decomposed in terms of a single hadronic form factor $V_i(z)$
\begin{align}
    \mathcal{A}_\mu^\text{i}  = & -i \, G_F \frac{V_i(z)}{(4\pi)^2} \left[ q^2 (k+p)_\mu - (M_K^2 - M_\pi^2) q_\mu \right] \,,
\end{align}
where $k$, $p$ and $q = k-p$ are the 4-momentum of the kaon, pion, and virtual photon respectively. The form factor is parameterised by the dimensionless ratio $z=q^2/M_K^2$, and has the general form $V_i(z) = a_i + b_i z + V^{\pi\pi}_i(z)$, where $V^{\pi\pi}$ contains the contribution from the intermediate $\pi\pi \to \gamma^\ast$ process \cite{DAmbrosio:1998gur}. 

The parameters $a_i$ and $b_i$ can be computed within the SM, as well as extracted from experiments, the values of which are given in \cref{tab:exp_theo_params}. It is clear that the experimental and theory values for $a_+$ are significantly discrepant, however, there are contributions to the theoretical value that have not currently been taken into account that are discussed in \cite{DAmbrosio:2019xph}. It is therefore important that the theory community work to produce a robust prediction for $a_+$ that includes all sources of statistical and systematic errors. Lattice QCD is the only method that allows us to compute such observables from first principles. For the rest of these proceedings we focus on the charged kaon decay mode, although all of the principles are also applicable to the neutral kaon mode.

\begin{table}[h!]
\centering
\begin{tabular}{c|cc|cc}
    & $a_+$ & $b_+$ & $a_S$ & $b_S$ \\
    \hline
    Exp. \cite{NA62:2022qes} \cite{NA481:2004nbc} & $-0.575(13)$ & $-0.722(43)$ & $-1.6^{+2.1}_{-1.8}$ & $10.8^{+5.4}_{-7.7}$ \\
    & & & \multicolumn{2}{c}{or} \\
    & & & $1.9^{+1.6}_{-2.4}$ & $-11.3^{+8.8}_{-4.5}$ \\
    \hline
    Theory \cite{DAmbrosio:2019xph} & $-1.59(8)$ & $-0.82(6)$ & - & -
\end{tabular}
\caption{Existing experimental and theory results for the form factor parameters $a_i$ and $b_i$ for the charged and neutral kaon decay channels.}
\label{tab:exp_theo_params}
\end{table}

\begin{figure}
    \centering
    \includegraphics[width=0.45\linewidth]{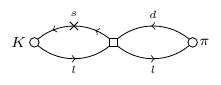}
    \includegraphics[width=0.45\linewidth]{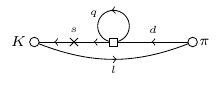}
    \includegraphics[width=0.45\linewidth]{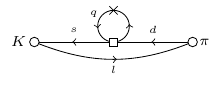}
    \caption{Representatives of the different classes of diagrams in the 4-point correlation function. These are the Non-Loop (top left), Loop (top right), Loop-Insertion (bottom) respectively. The Non-Loop and Loop topologies also have additional diagrams with the current inserted on different legs. There are additional disconnected diagrams that are not considered in this study.}
    \label{fig:RK_diagrams}
\end{figure}

\subsection{Existing lattice result}
The framework for computing this decay process on the lattice was introduced in \cite{Christ:2015aha}, the key object of which is the Euclidean space-time 4-point correlation function
\begin{align}
\label{eq:4ptCorr}
    \Gamma_\mu^i(t_K,t_H,t_\pi) = \langle \phi_{\pi^i}(t_\pi, \boldsymbol{p}) \, H_W(t_H, \boldsymbol{0}) \, J_\mu(0,\boldsymbol{q}) \, \phi_{K^i}^\dagger(t_K, \boldsymbol{k}) \rangle \,,
\end{align}
where all operators are in the time-momentum representation, and $\phi_{\pi^i}$ and $\phi_{K^i}$ are interpolating operators with the quantum numbers of the the pion and kaon respectively. There are several types of Wick contraction that go into this 4-point function, representatives of which are shown in \cref{fig:RK_diagrams}, which we denote as Non-Loop (NL), Loop (L), and Loop-Insertion (LI) diagrams. There are additional diagrams where the vector current is located on a disconnected loop that are neglected due to them being color and $\text{SU}(3)$ flavor suppressed. 

The connection between the correlator in \cref{eq:4ptCorr} and the physical amplitude in \cref{eq:Amplitude_Def} is complicated by the breakdown of the analytic continuation, which manifests as exponentially growing terms from intermediate states that must be removed, the details of which can be found in \cite{Christ:2015aha}. For the purposes of these proceedings, all that is required is that the objects that most closely relate to the amplitude are the integrated correlators
\begin{align}
    I^\rho_\mu(T_a) = -i \int_{-T_a}^{0} dt_H \Gamma_\mu(t_H) \hspace{2em}\text{and}\hspace{2em} I^\sigma_\mu(T_b) = -i \int_{0}^{T_b} dt_H \Gamma_\mu(t_H) \,,
\end{align}
which separates the two time orderings of the intermediate operators. In \cite{Christ:2015aha,Christ:2016mmq,RBC:2022ddw} they alternatively use $I_\mu(T_a,T_b) = I^\rho_\mu(T_a) + I^\sigma_\mu(T_b)$ which contains the same information.

The first lattice calculation of this decay performed at the physical point was presented in \cite{RBC:2022ddw}, with the final result
\begin{align}
\label{eq:lat_value}
    a_+ \simeq V_+(z=0.013) = -0.87(4.44) \,,
\end{align}
which has an uncertainty approximately $8\times$ larger than the experimental central value and $3\times$ the theory value. Therefore up to an order of magnitude error reduction is required to be sensitive to the discrepancy between the theory and experimental values. 
The large uncertainty in \cref{eq:lat_value} is due to the stochastic estimation of loop propagators required in the L and LI diagrams which contain the difference between a light and a charm loop resulting in a large cancellation. At unphysically large light quark mass \cite{Christ:2016mmq} this cancellation was not problematic due to a large correlation between the two loops, therefore a reasonable signal could could be achieved despite the cancellation. However, at physical quark masses, this correlation is significantly reduced resulting in serious degradation of the final signal. Achieving the required error reduction using the same techniques as \cite{RBC:2022ddw} would require an unreasonably large increase in computational resources, it is therefore imperative that an improved method for calculating the problematic loop differences is found.

\section{Split-even approach}
The object required in the computation of the L diagrams is
\begin{align}
    \Delta L(x) = D_l^{-1}(x|x) - D_c^{-1}(x|x) \,,
\end{align}
where $D_q$ is a discretisation of the Dirac operator of quark flavor $q$. In order to compute this exactly, one would have to invert the Dirac operator for each lattice site $x$ which would be prohibitively expensive. Instead it is common practice to estimate it stochastically by introducing a set of noise fields $\eta_i(x)$ that satisfy
\begin{align}
    \lim_{N_s \to \infty} \frac{1}{N_s} \sum_{i=1}^{N_s} \eta_i(x) = 0 \hspace{2em} \text{and} \hspace{2em} \lim_{N_s \to \infty} \frac{1}{N_s} \sum_{i=1}^{N_s} \eta_i(x) \eta_i^\dagger(y) = \delta(x-y) \,.
\end{align}
The loop estimator is therefore given by
\begin{align}
    L_q^\text{std}(x) = \frac{1}{N_s}\sum_{i=1}^{N_s} D_q^{-1}(x|\eta_i) \eta_i^\dagger(x) \, \overset{N_s \to \infty}{\to} D^{-1}_q(x|x) \,,
\end{align}
where we have defined the shorthand notation $D^{-1}(x|\eta) = \sum_y D^{-1}(x|y) \eta(y)$. The loop difference is simply given by the difference of these, $\Delta L^\text{std}(x) = L_l^\text{std}(x) - L_c^\text{std}(x)$. This is exactly the estimator used in the previous calculation where it was observed a large correlation between the two terms is required for a reasonable signal to be obtained \cite{RBC:2022ddw}. We refer to this approach as the standard estimator.

The split-even estimator proposed in \cite{Giusti:2019kff} avoids such issues with cancellations by directly computing the difference. It utilises the identity
\begin{align}
    D_l^{-1} - D_c^{-1} = (m_c-m_l) D_l^{-1} D_c^{-1} \,,
\end{align}
which is true for Wilson fermions as well as for Domain-Wall fermions (see \cite{Harris:2023zsl} for details).
After applying this identity, the noise fields can be inserted at the rightmost position, which is then identical to the standard estimator, or alternatively in-between the two inverse Dirac operators giving the split-even estimator
\begin{align}
    \Delta L^\text{split}(x) = (m_c-m_l) \frac{1}{N_s}\sum_{i=1}^{N_s} D_l^{-1}(x|\eta_i) D_c^{-1}(\eta_i^\dagger|x) \,.
\end{align}
Note that this is simply a different combination of the same propagators needed for the standard estimator, and therefore has the same computational cost.

The split-even estimator can be applied trivially to the L diagrams of the 4-point correlation function, because they contain exactly such loop differences. However, the LI diagrams do not immediately contain the difference of propagators required, but instead contains a difference of products of propagators (separated by a vector current)
\begin{align}
    \Delta L_\mu(x) = [D_l^{-1} \mathcal{V}_\mu D_l^{-1}](x|x) - [D_c^{-1} \mathcal{V}_\mu D_c^{-1}](x|x) \,,
\end{align}
where $\mathcal{V}_\mu$ is a flavor-singlet vector current kernel, which may be either a local or conserved current, and may be restricted to a single time plane and projected to some non-zero momentum transfer, as is the case in this study.
This can however be transformed into a form in which the split-even estimator is applicable by adding and subtracting a flavor-changing vector current term $D_l^{-1} \mathcal{V}_\mu D_c^{-1}$
\begin{align}
    \Delta L_\mu(x) = & [D_l^{-1} \mathcal{V}_\mu D_l^{-1}](x|x) - [D_l^{-1} \mathcal{V}_\mu D_c^{-1}](x|x) \\
    + & [D_l^{-1} \mathcal{V}_\mu D_c^{-1}](x|x) - [D_c^{-1} \mathcal{V}_\mu D_c^{-1}](x|x) \nonumber \\
    = & (m_c-m_l) \left( [D_l^{-1} \mathcal{V}_\mu D_l^{-1} D_c^{-1}](x|x) + [D_l^{-1} D_c^{-1} \mathcal{V}_\mu D_c^{-1}](x|x) \right) \,.
\end{align}
This therefore gives the sum of two terms which each have a product $D_l^{-1} D_c^{-1}$ between which the noise fields can be inserted to give the split-even estimator of the LI diagrams as
\begin{align}
    \Delta L_\mu^\text{split}(x) = (m_c-m_l) \frac{1}{N_s} \sum_{i=1}^{N_s} & \left( [D_l^{-1} \mathcal{V}_\mu D_l^{-1}](x|\eta_i) D_c^{-1}(\eta_i^\dagger|x) \right. \\
    & \left. + D_l^{-1}(x|\eta_i) [D_c^{-1} \mathcal{V}_\mu D_c^{-1}](\eta_i^\dagger|x) \right) \,. \nonumber
\end{align}

In addition to the use of the split-even estimator, the loop difference can be broken up into multiple smaller differences with intermediate mass quarks $l-c = (l-c_1)+(c_1-c_2)+...+(c_{N}-c)$, each of which can utilise the split-even estimator. This technique, known as frequency-splitting, allows the computational effort to be distributed according to the regions of the mass spectrum that contribute most to the statistical uncertainty. It does however require Dirac operator inversions for the additional intermediate quarks (a trade-off that must be balanced with the reduced number of hits required for the smaller differences).

\section{Numerical results}
In this study we use the same RBC-UKQCD physical point ensemble \cite{RBC:2014ntl} that was used in the previous rare kaon decay calculation \cite{RBC:2022ddw}. That is a $2+1$ flavor ensemble with physical light and strange quark masses and an inverse lattice spacing of $a^{-1} = 1730 \, \text{MeV}$. All sea quarks and the strange valance quark are simulated using the M\"obius Domain-Wall action, while the light and charm valance quarks are simulated with the zM\"obius action \cite{Mcglynn:2015uwh} tuned to approximate the M\"obius action with smaller $L_s$. This introduces a bias which should be corrected with an All-Mode-Averaging (AMA) type procedure \cite{Blum:2012uh}. However, this correction is expected to be small due to the tuning of the zM\"obius action so will be neglected in this study.

It has been observed that the Domain-Wall formalism breaks down for sufficiently heavy quark masses \cite{Boyle:2016imm}, and therefore the physical charm quark cannot be simulated directly at this lattice spacing. We therefore make measurements with the 3 lighter-than-physical charm masses used in \cite{RBC:2022ddw} for an extrapolation to physical charm mass. This combines very naturally with the frequency-splitting technique where the intermediate charm masses can be those needed for the extrapolation, and therefore the frequency-splitting does not require additional propagators for those masses. We also include 2 additional lighter intermediate masses, the lightest of which is equal to the strange quark, to help isolate which regions of the mass spectrum contribute most to the noise. The bare charm masses used and the corresponding $\eta_c$ masses are shown in \cref{tab:charm_masses}.
\begin{table}[]
    \centering
    \begin{tabular}{c|c|c|c|c|c}
        & $c_1$ & $c_2$ & $c_3$ & $c_4$ & $c_5$  \\
        \hline
       $am_c$ & $0.0362$ & $0.15$ & $0.25$ & $0.30$ & $0.35$ \\
       $M_{\eta_c}$ [MeV] & 693(2) & 1492(1) & 2007(1) & 2230(1) & 2432(1)
    \end{tabular}
    \caption{Bare quark masses (in lattice units) and corresponding $\eta_c$ meson masses for the 5 intermediate charm quarks used in this study.}
    \label{tab:charm_masses}
\end{table}

Since the results presented here are only an exploratory study of the split-even method for this process, we use low statistics with only 10 configurations and average over 6 translations in time. For the stochastic noises we use $N_s = 32$ hits for all mass splittings.

\begin{figure}
    \centering
    \includegraphics[width=0.45\linewidth]{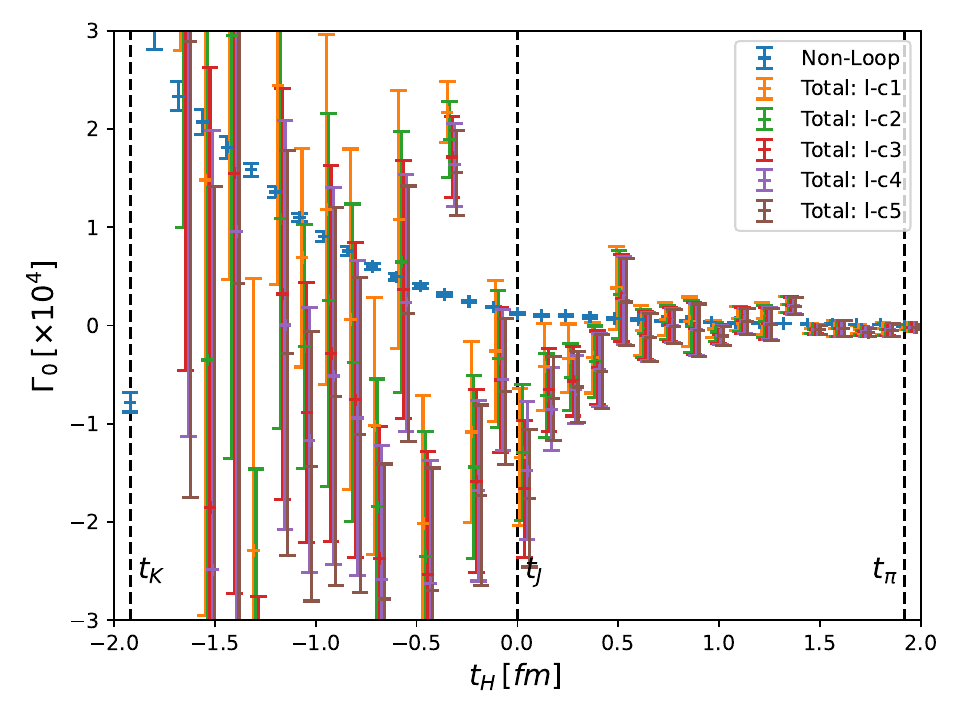}
    \includegraphics[width=0.45\linewidth]{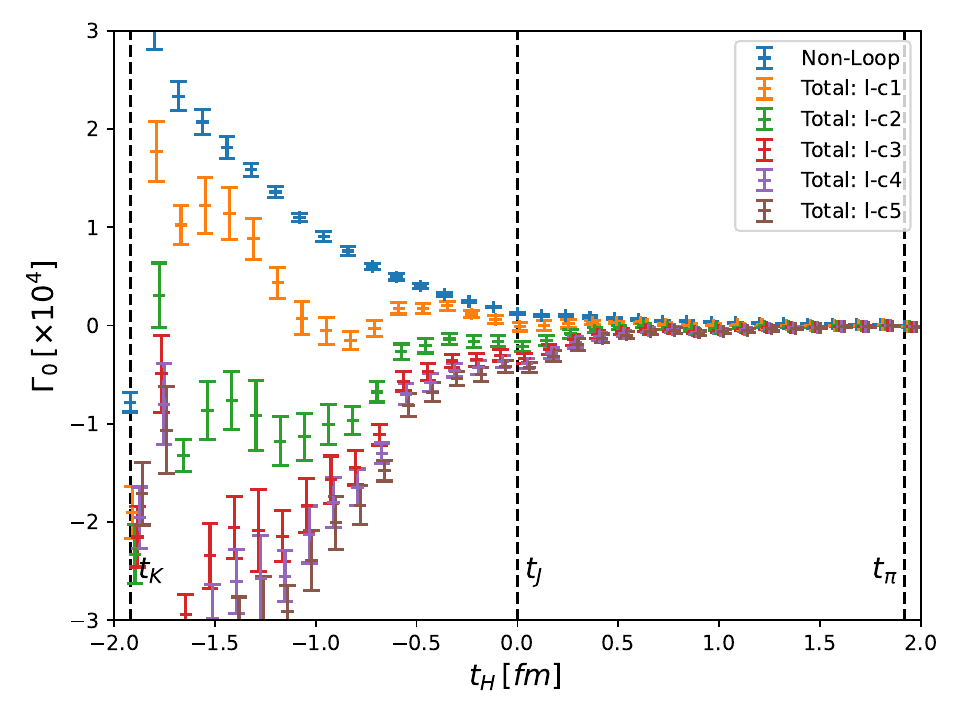}
    \caption{Comparison between the 4-point correlation function in \cref{eq:4ptCorr} for different charm masses using the standard (left) and split-even (right) estimators. The non-loop diagram is shown for reference.}
    \label{fig:corr_cmp}
\end{figure}

\Cref{fig:corr_cmp} shows a comparison of the 4-point correlation in \cref{eq:4ptCorr} using the standard and the split-even estimators at equal computational cost. It is clear that the split-even estimator provides a large reduction in the uncertainties, which range from an approximately $5-25 \times$ error reduction.

\begin{figure}
    \centering
    \includegraphics[width=0.45\linewidth]{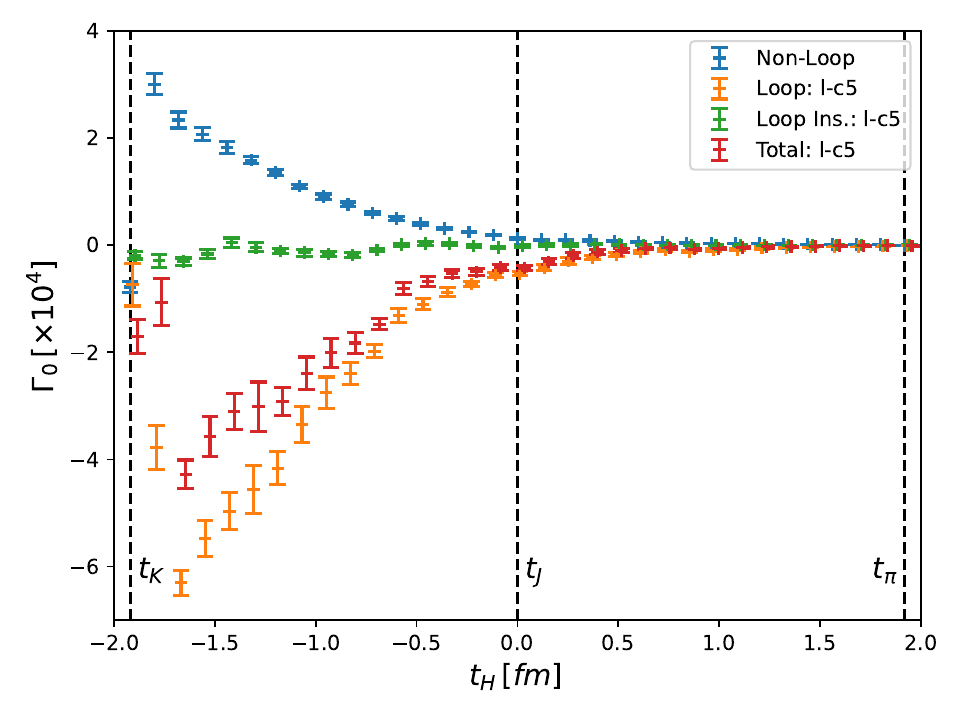}
    \includegraphics[width=0.45\linewidth]{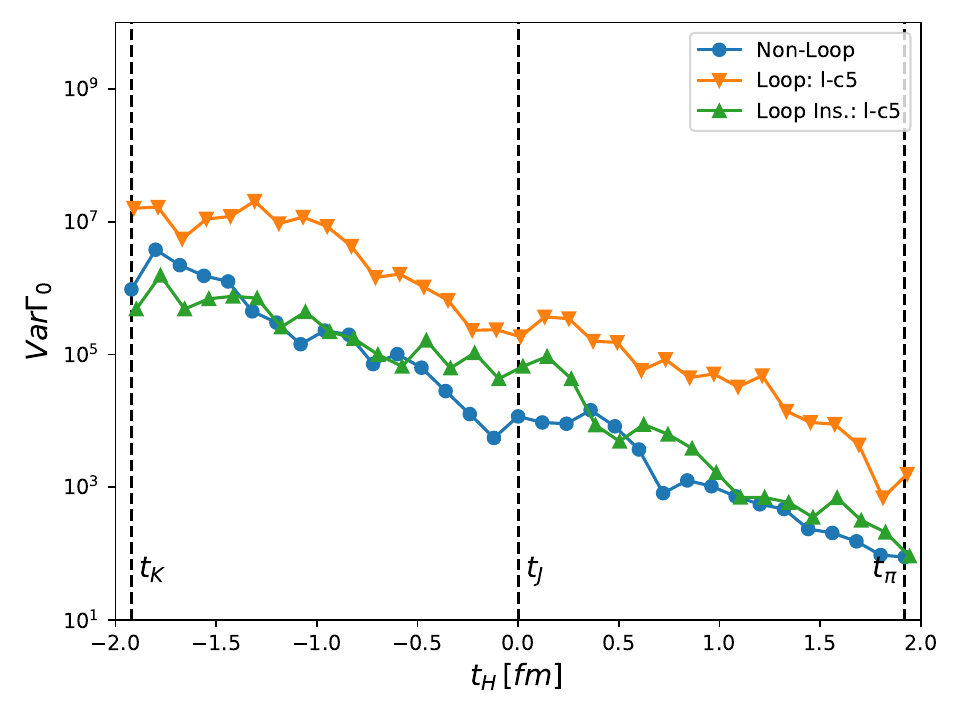}
    \caption{The 4-point correlation function (left) and the corresponding variance (right), broken down into the NL, L, and LI diagrams.}
    \label{fig:corr_var_breakdown}
\end{figure}

\Cref{fig:corr_var_breakdown} shows a breakdown of the contributions from the different types of diagrams. We observe that the LI diagrams contribute only a very small amount to to total, as well as contributing to the variance at a level below that of the L diagrams. Since the LI diagrams are the most expensive to compute, we may therefore benefit greatly by reducing the computational effort spent on these diagrams, for example by performing fewer time translations and/or reducing the number of noise hits.

\begin{figure}
    \centering
    \includegraphics[width=0.45\linewidth]{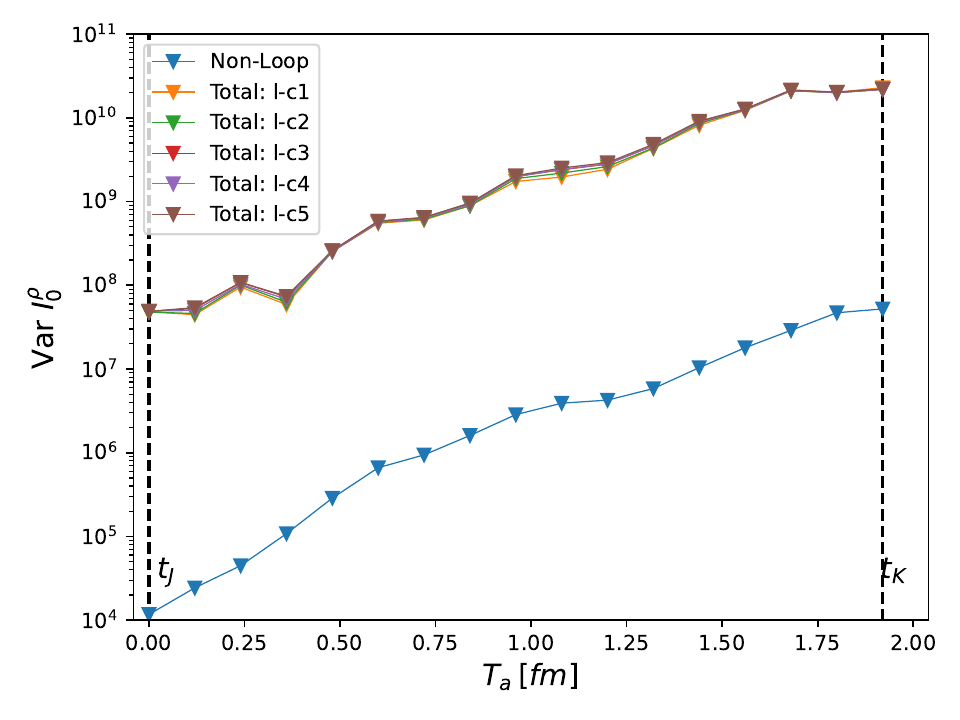}
    \includegraphics[width=0.45\linewidth]{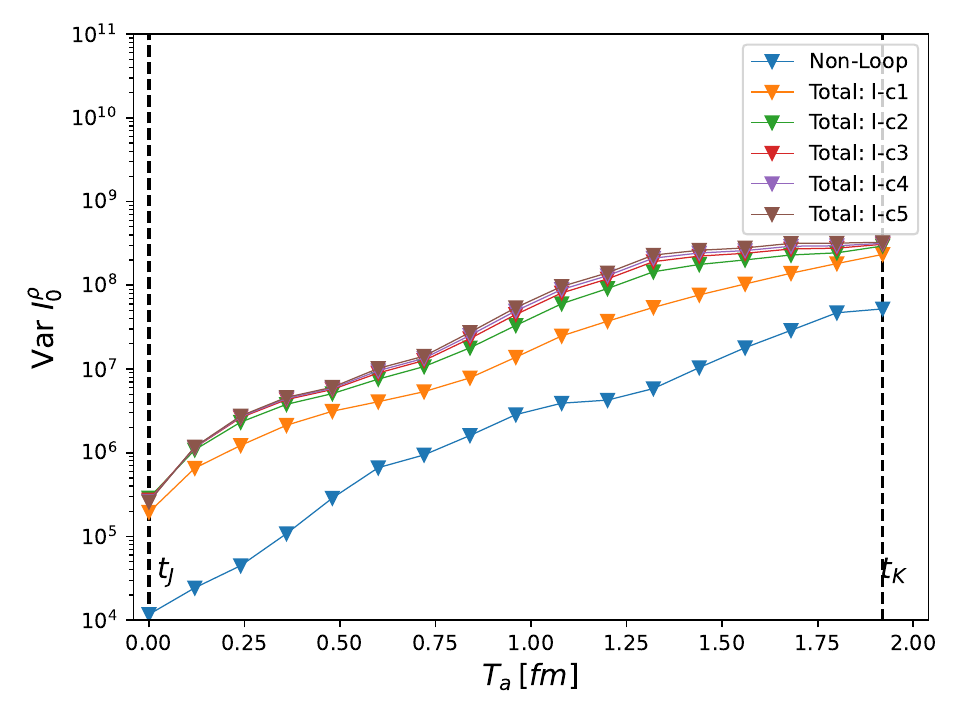}
    \includegraphics[width=0.45\linewidth]{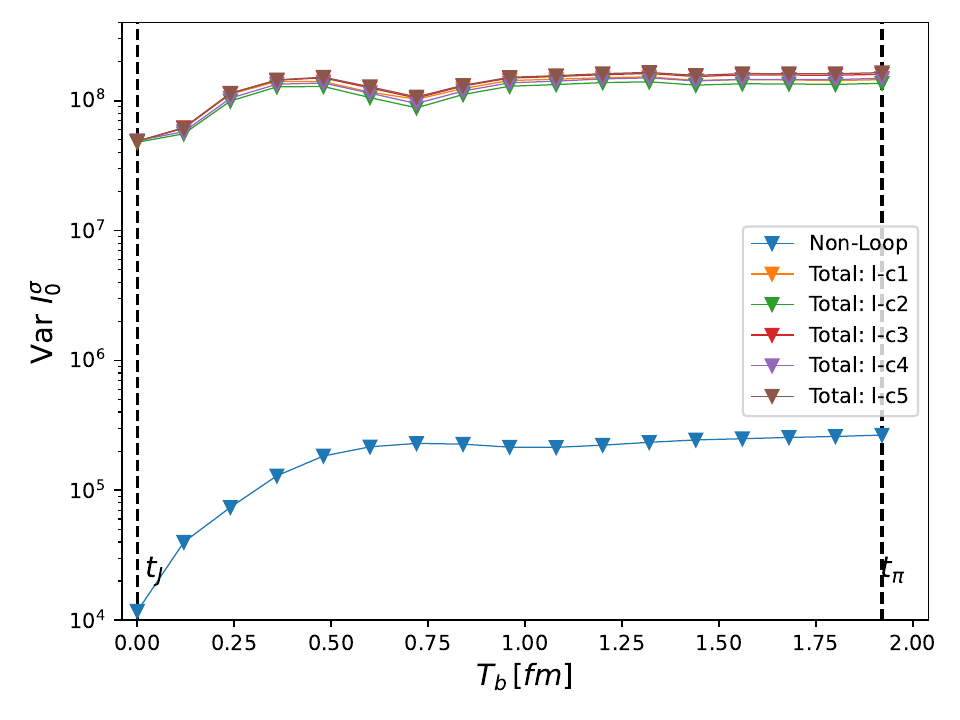}
    \includegraphics[width=0.45\linewidth]{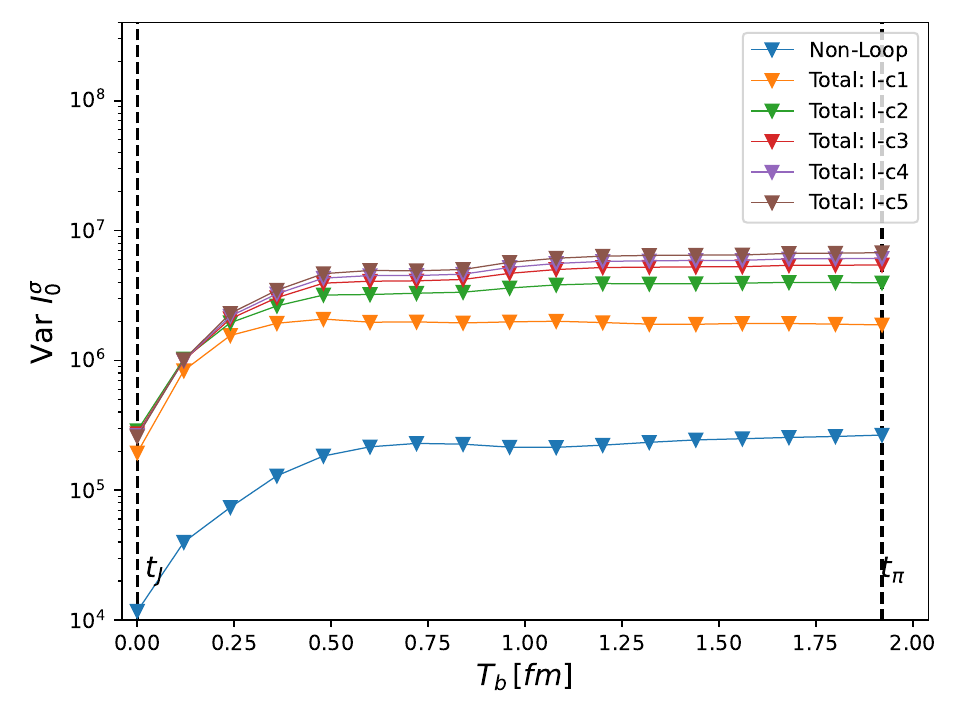}
    \caption{Variance of the two integrated correlators (top and bottom) using the standard (left) and split-even (right) estimators.}
    \label{fig:int_var_cmp}
\end{figure}

In \cref{fig:int_var_cmp} we see the comparison of the variance of the integrated correlators between the two methods, and for each of the different charm masses. It is clear that much of the statistical improvements seen in figure \cref{fig:corr_cmp} have propagated into the integrated correlator, with an error reduction of between $4-10\times$ for different integrals and charm masses. We also see that much of the remaining noise comes from the lightest mass difference $l-c_1$. Therefore the frequency-splitting technique can be utilised to perform additional hits in the light part of the spectrum, while spending less on the heavier charm masses that we see contribute very little to the total error. 

\section{Conclusions and Outlook}
In conclusion, we have demonstrated that the split-even estimator proposed in \cite{Giusti:2019kff} is applicable to the Loop and Loop-Insertion diagrams in the lattice calculation of the rare kaon decay $K^+ \to \pi^+ \ell^+ \ell^-$, which are exactly the dominant source of uncertainty in the existing calculation of this process \cite{RBC:2022ddw}. We have seen that this improved estimator has shown a massive reduction in the statistical error of the 4-point correlation function and consequently in the integrated 4-point functions that would be used to extract the amplitude in a future analysis. Assuming these improvements propagate to the final form factor, they are at the level required in order to be sensitive to the discrepancy between existing theory and experimental results. In addition it is observed using the frequency-splitting technique that the largest contributor to the remaining uncertainty is in the light part of the spectrum, which suggests more computational effort should be spent there and less on the heavy mass differences.

This being an exploratory study of the methodology, there are limitations that must be lifted, such as performing the zM\"obius-to-M\"obius bias correction, and the inclusion of disconnected diagrams to which the split-even estimator can also be applied \cite{Giusti:2019kff}. Also due to the low statistics nature of this study study, measurements will need to have to be made on additional configurations, along with additional noise hits.

While it is expected that these error reductions on the integrated correlator propagate through the analysis into the final value of the form factor, due to the relative complexity of such an analysis it is not certain if all of the improvement will be realised, or if new statistical or systematic limiting factors comes to dominate the error. This is all to be investigated in future work.

\acknowledgments
This work used the DiRAC Extreme Scaling service Tursa at the University of Edinburgh, managed by the Edinburgh Parallel Computing Centre on behalf of the STFC DiRAC HPC Facility (www.dirac.ac.uk). The DiRAC service at Edinburgh was funded by BEIS, UKRI and STFC capital funding and STFC operations grants. DiRAC is part of the UKRI Digital Research Infrastructure.
V.G. and A.P. are supported in part by UK STFC grants ST/X000494/1 \& ST/T000600/1. 
R. Hill is supported by UK STFC grant ST/T000600/1.
A.P. also received funding from the European Research Council (ERC) under the European Union’s Horizon 2020 research and innovation programme under grant agreements No 757646 \& 813942.

\newpage

\bibliographystyle{JHEP}
\bibliography{refs}

\end{document}